\documentclass[12pt,preprint]{aastex}

\shorttitle{A 600 min NIR lightcurve of Sgr~A*}
\shortauthors{Meyer et al.}

\begin{document}


\title{A 600 minute near-infrared lightcurve \\
    of Sagittarius A*}


\author{L. Meyer\altaffilmark{1}, T. Do, A. Ghez, M. R. Morris}
\affil{Department of Physics and Astronomy, University of California,
    Los Angeles, CA 90095-1547}

\author{G. Witzel, A. Eckart}
\affil{Universit\"at zu K\"oln, Z\"ulpicher Str. 77, 50937 K\"oln}

\author{G. B\'{e}langer}
\affil{ESA/ESAC, PO Box 78, 28691 Villanueva de la Ca\~{n}ada, Spain}

\author{R. Sch\"odel}
\affil{Instituto de Astrof\'{i}sica de Andaluc\'{i}a, Camino Bajo de Hu\'{e}tor 50, 18008 Granada, Spain}


\altaffiltext{1}{Supported by a fellowship within the Postdoc-Program of the German Academic Exchange Service (DAAD).}

\begin{abstract}
We present the longest, by a factor of two, near-infrared lightcurve from Sgr~A* -- the supermassive black hole in the Galactic center. Achieved by combining Keck and VLT data from one common night, which fortuitously had simultaneous Chandra and SMA data, this lightcurve is used to address  two outstanding problems.  First, a putative quasi-periodicity of $\sim$20\,min reported by groups using ESO's VLT is not confirmed by Keck observations.   Second, while the infrared and  mm-regimes  are thought to be related based on reported time lags between lightcurves from the two wavelength domains, the reported time lag of 20\,min inferred using the Keck data of this common VLT/Keck night only is at odds with the lag of $\sim100$\,min reported earlier.  With our long lightcurve, we find that (i) the simultaneous 1.3 millimeter observations are in fact consistent with a $\sim100$\,min time lag, (ii) the different methods of NIR photometry used by the VLT and Keck groups lead to consistent results, (iii) the Lomb-Scargle periodogram of the whole NIR lightcurve is featureless and follows a power-law with slope -1.6, and (iv) scanning the lightcurve with a sliding window to look for a transient QPO phenomenon reveals for a certain part of the lightcurve a 25\,min peak in the periodogram. Using Monte Carlo simulations and taking the number of trials into account, we find it to be insignificant.   
\end{abstract}


%\facilities{VLT, Keck, SMA}



\section{Introduction}

The near-infrared (NIR) regime at high angular resolution has proven to be of great value in Galactic center research. The proper motion of stars detected in this waveband demonstrates the existence of a supermassive black hole (BH) at the center of our Galaxy: Sagittarius~A* \citep[Sgr~A*; see, e.g.,][]{eckigenzel, ghez98, gheznature,ghez,genzel00,rainer1}. In 2003, NIR emission associated with Sgr~A* was detected \citep{genzel,ghez04}, which is important since it is the most underluminous BH accretion system observed thus far (with a bolometric luminosity nine orders of magnitude lower than its Eddington luminosity). 
The NIR emission is highly variable: intensity changes by factors $\leq 10$, lasting between about 10 and 100 min, occur at least 4 times a day, and are highly polarized \citep[e.g.][]{ecki1,ecki2,ich2,ich,tuan,trippe}. Simultaneous NIR and X-ray observations have revealed that each X-ray flare is accompanied by a NIR flare with zero time lag, but not vice versa \citep{ecki04,ecki1,ecki08,belanger05,hornstein07}. 
Recent campaigns that included mm observations, reported a characteristic time lag between the NIR/X-ray and longer wavelengths that has been interpreted in terms of an expanding synchrotron plasmon. The limited overlap between data sets, however, has led to debates over the nature of this time lag \citep[e.g.][]{ecki1,yusef06,yusef07,danmultwav}.

The characteristics of the NIR emission from Sgr~A* is of particular interest given that there have been reports of periodic modulations with a period of $\sim$20\,min present in NIR flares detected during VLT observations \citep{genzel,ecki2,ich2,ich,trippe}. In a recent study, however, \citet{tuan} used robust statistical estimators and found no significant peaks in the periodograms of the Sgr~A* lightcurves observed with the Keck II telescope.  Moreover, quasi-periodicities in the X-ray regime claimed to be present by \citet{aschenbach1}, are not statistically significant \citep{belanger}.

In this Letter, we address the questions of a periodic component in the NIR flux, and the relation of the NIR to the mm-regime, by combining for the first time contiguous Keck and VLT data. During that 10 hour session, there was also simultaneous coverage by Chandra and SMA. These observations were published by \citet{danmultwav}, but interpreted taking only the Keck data into account.

\section{The data}

During the night of 2005 July 30-31, the VLT observed Sgr~A* from 23:05\,UT to 06:53\,UT using the Natural Guide Star Adaptive Optics system and NIR camera NACO on UT4\footnote{ESO program 075.B-0093(B).}. Since these observations have not been published before, the observational details are given here.
The detector integration time was 15\,s, and four images were co-added before the data were recorded: the effective resolution is $\sim$1 image/80\,s. Dithering was used to minimize the effects of dead pixels. The atmospheric seeing conditions ranged between $1\arcsec-2.5\arcsec$ (as determined by the differential image motion monitor) during the first two hours, and $0.5\arcsec-1.25\arcsec$ afterwards (the average Strehl ratio of the VLT data is 17\%). 

At 07:00\,UT, the Keck II telescope started its monitoring campaign using the NIRC2 camera in combination with Laser Guide Star AO. The Keck observations were made by cycling through the H-K-L wavelength filters \citep[see][for details]{hornstein07}. Since the VLT observations were carried out in K-band, we consider only the K-band subset of the Keck data.

Both data sets were reduced in the same standard way, i.e. sky subtracted, flat-fielded, and corrected for bad pixels. Images with a  Strehl ratio less than 10\% were removed (13 out of 266 VLT images). For every individual image, the point spread function (PSF) was extracted with the code StarFinder by \citet{diolaiti}. Each exposure was deconvolved with a Lucy-Richard deconvolution and restored with a Gaussian beam. 
The flux of Sgr A* and other compact sources in the field were obtained via aperture photometry on the diffraction limited images with a circular aperture of radius $0\arcsec.03$. The background flux density was determined as the mean flux measured with apertures of the same size at five different positions in a field located $1\arcsec$ northwest of Sgr~A* that shows no individual stars. Photometric calibration was done relative to stars in the field with known flux. For the extinction correction we assumed $A_K = 2.8$\,mag \citep{eisendings05}. Estimates of uncertainties were obtained from the standard deviation of fluxes of nearby constant sources.

It is noteworthy that this is the first time that Galactic Center VLT and Keck data have been reduced homogeneously. While the VLT groups mainly use deconvolution and aperture photometry, the Keck group uses PSF fitting without deconvolution. 
The Keck part of the lightcurve presented in the next section is similar to that reported by \citet{hornstein07}, where the PSF fitting technique was applied. After rescaling the lightcurve by a multiplicative factor to account for a different de-reddening factor and slightly different calibration values used by \citet{hornstein07}, the average difference per data point between both methods is only 0.031\,mJy (de-reddened), well within our $1\sigma$-error bar of 0.18\,mJy.  Both data reduction methods are consistent with each other, and no large systematic errors are introduced by choosing either one.

\section{Results}
\subsection{The NIR properties of Sgr A*}

\begin{figure}
%\epsscale{.50}
%\plottwo{31_07_2005_nolines.eps}{31-Jul-05_Perio_Sgr.eps}
\begin{minipage}{8.5cm}
\includegraphics[angle=270,scale=.35]{f1a.eps}
\end{minipage}
\begin{minipage}{8.5cm}
\includegraphics[scale=.5]{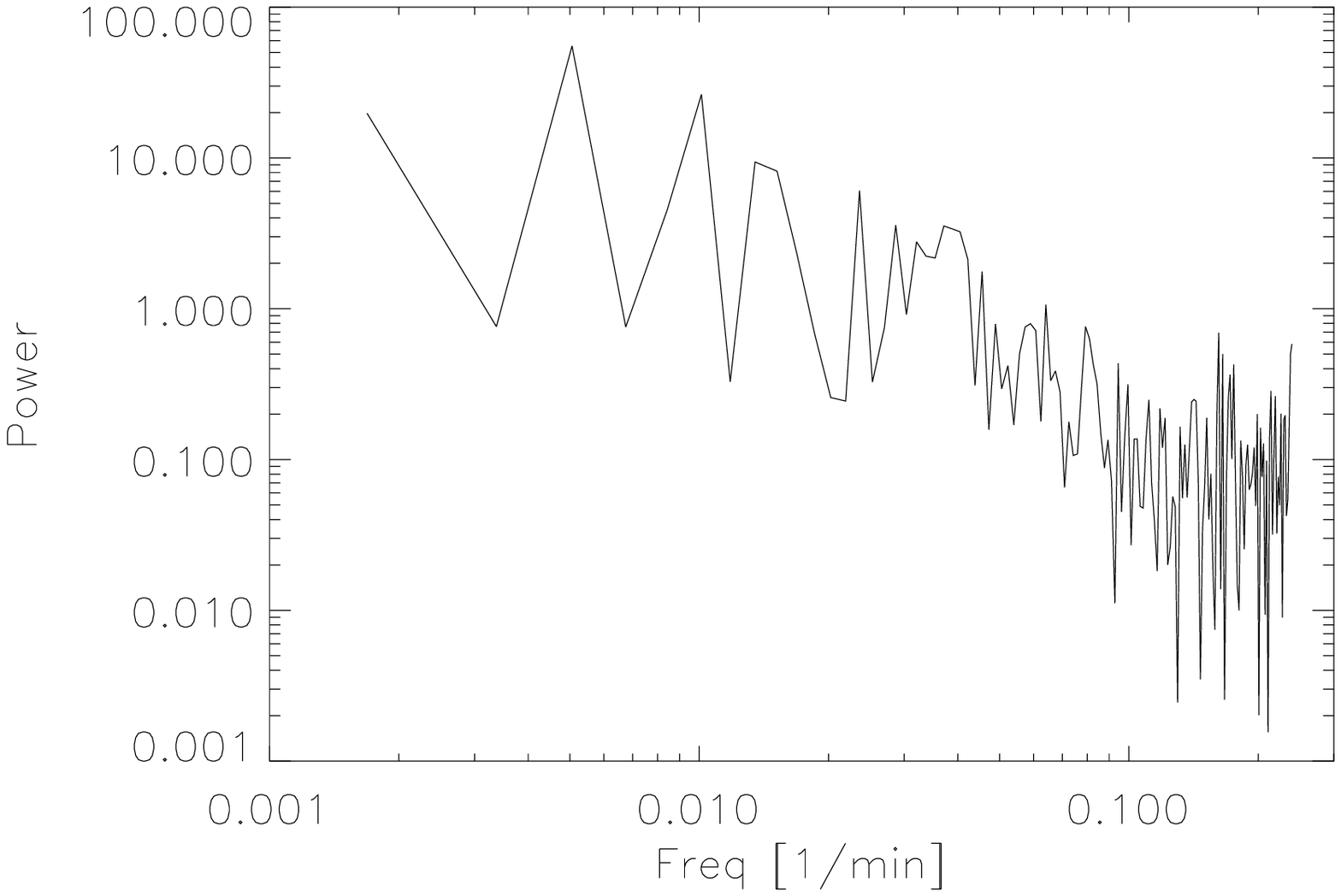}
\end{minipage}
\caption{Left panel: De-reddened lightcurve from Sgr A* (lower curve) on the night of 2005 July 30-31 (UT) and from a comparison star (S0-8 shifted 5\,mJy upwards; upper curve). The circles are the VLT data points (0-450\,min) and the triangles are the Keck data points (450-580\,min). The small gaps in the lightcurve are due to either periods of very bad seeing leading to useless data, sky observations, or laser collision at Keck. Right panel: Lomb-Scargle periodogram of the lightcurve. No smoothing was applied.}
\label{fig1}
\end{figure}

Figure~\ref{fig1} (left panel) shows the 2005 July 30-31 de-reddened lightcurve. The first 450 min are the VLT data, and the following 130 min are the Keck data. Note that the error bars are smaller for the Keck data due to better seeing conditions at Mauna Kea that night. The corresponding Lomb-Scargle periodogram for the entire lightcurve is presented in the right panel of Fig.~\ref{fig1}. It is consistent with the finding of \citet{tuan} that the NIR emission of Sgr~A* is described by a single stochastic process (other than additive measurement noise) that has a power-law spectrum, $\mathcal{P}\propto f^{-\alpha}$, very similar to the X-ray emission of AGN. The Lomb-Scargle periodogram in Fig.~\ref{fig1} is a single realization of this process and therefore fluctuates around the spectrum. We determine the probability density function (PDF) of these fluctuations around the spectrum at a given frequency empirically with Monte Carlo simulations. With the PDF at hand, we can assess the likelihood that peaks in the periodogram are not due to a fluctuation but rather have a physical cause intrinsic to the source. 

We used two different Monte Carlo based analyses recently developed by \citet{tuan} and \citet{belanger} to look for significant peaks in the periodogram. These analyses are carried out by first determining the power-law index of the spectrum of the stochastic process, and then generating lightcurves \citep[realizations of this process;][]{timmer95} with  a sampling function matching those of the data set. Finally, the significance of each observed periodogram peak is derived from the large simulated reference data set.  

An accurate determination of the power-law index $\alpha$ of the underlying stochastic process is important in this approach. \citet{belanger} do this by first performing several estimates of $\alpha$ by fitting a power-law to the periodogram made from the lightcurve binned with successively larger bin times, and then carrying out simulations with the same count rate and sampling to find the matching curve of $\alpha$ versus bin time. \citet{tuan} also use MC simulations, but determine the power-law index using the structure function, which also takes the sampling into account. 
Both methods lead to a spectral index of $\alpha=1.6\pm 0.05$ (formal fitting error). Note that the length of the lightcurve allows to sample frequencies $< 10^{-4}$\,Hz ($0.006\,\mbox{min}^{-1}$) for the first time, showing that the power-law extends to this regime. Unfortunately, we cannot fully exclude the possibility of spectral leakage from a process with a spectral index steeper than -1.6. However, our main conclusions in this paper also hold true for steeper power-law indices. We plan to use more sophisticated spectral estimators elsewhere.

The periodogram in the right panel of Fig.~\ref{fig1} clearly shows no outlying peaks above the underlying spectrum of the power-law process.
However, the featurelessness of the periodogram of the whole lightcurve does not rule out the presence of a periodic component in parts of it, as the mechanism giving rise to such a component may be transient and short lived. We therefore did a search for periodicities over a restricted range of frequencies by scanning the lightcurve using a sliding window method where window here means a sub-span of the time series. This consists of constructing a periodogram for the data subset corresponding to each window, and assigning a significance to each point based on the probability density functions derived from the simulations of the whole lightcurve described above.

\begin{figure}
%\begin{minipage}{8.5cm}
\includegraphics[scale=.65]{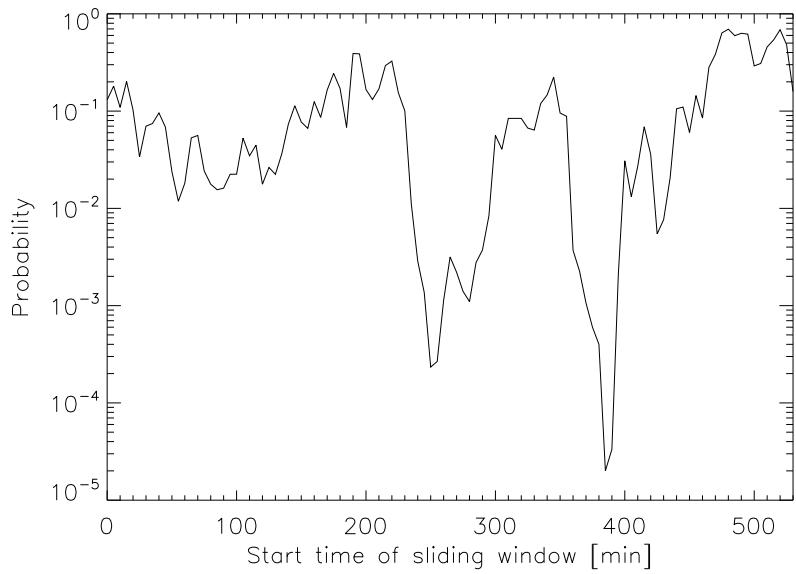}
\caption{Sliding window scan of the lightcurve seen in Fig.~\ref{fig1}. Each time window has a length of 60\,min and has been shifted by 5\,min for the scan. The abscissa shows the start time of each sliding window. The ordinate shows the probability that the most significant peak in the corresponding periodogram is due to a fluctuation around the power-law spectrum of the red noise stochastic process. A very low probability means that an additional  \mbox{(quasi-)periodic} component is needed to explain the periodogram peak. See text for more details.}
\label{scan}
%\end{minipage}
\end{figure}

Fig.~\ref{scan} shows the result of such a scan using a 60\,min window with 5\,min steps. The probability that the most significant peak in each periodogram is due to a statistical fluctuation around the power-law spectrum is plotted against the start time of the sliding window. The most significant peak overall (the one with the lowest probability to be due to a fluctuation) is found in the window starting at minute 385, and we thus looked at this window subset in more detail. 
Indeed, the flux between 385 and 445 min looks very similar to the 'sub-flare' phenomenology reported by \citet{genzel}, \citet{ecki1,ecki2}, and \citet{ich2}. These sub-flares are flux peaks superimposed on broader, longer lasting flux excursions and are thought to be the manifestation of the claimed quasi-periodicity. \citet{ich2,ich} and \citet{ecki08} showed that they can be interpreted in terms of a relativistically orbiting spot whose emission adds to the emission of the accretion flow. The NIR flux can therefore be described as $F(t)=A(t)+M(t)+S(t)$ with $A(t)$ a stochastic process with a power-law spectrum as above, $M(t)$ uncorrelated measurement noise, and $S(t)$ the deterministic flux of an (evolving) orbiting spot leading to a (quasi-)periodicity. Our MC simulations include $A(t)$ and $M(t)$ so that a possible quasi-periodic component can be identified as significant in the periodogram.

Fig.~\ref{scan} shows that there is a peak in the periodogram corresponding to the $385 - 445$\,min subset which has a false alarm probability of only $2\cdot 10^{-5}$ (this corresponds to $4.2\sigma$ in Gaussian equivalent terms, but note that the PDF is not Gaussian). This peak occurs at a frequency of $0.04\, \mbox{min}^{-1}$ (25\,min). However, we have to ask the question how likely it is that a periodogram peak with such a low probability does not only occur at this window, but in \textit{any} window. After all, there is \textit{a priori} nothing special about this certain sub-span of the lightcurve. We therefore count how often a peak with probability $\leq 2\cdot 10^{-5}$ occurs in our simulations in any window: for a fixed window length of 60\,min, we find 3120 occurrences for 30,000 simulations. This implies that the overall false alarm probability of the peak in the $385 - 445$\,min subset is $0.104$ (corresponding to $1.6\sigma$). Furthermore, we have to take into account the trials with windows of different lengths in scanning the data: this immediately yields a final significance of $\leq 1\sigma$.

Hence, we conclude that the whole lightcurve is consistent with a pure power-law process and no periodic component is needed. It is important to point out, however, that if we had observed (or analyzed) only the part of the data between $385 - 445$\,min, our result would have been interpreted as a $4\sigma$ detection of a 25\,min QPO. This points to an explanation of the contradictory results of \citet{genzel}, \citet{ecki1,ecki2} and \citet{ich2}, on the one hand, and \citet{tuan}, on the other. Our results here seem to favor the finding of \citet{tuan} where no periodicity was detected. 

It is interesting to note that the 25\,min periodogram peak is most significant for a 60\,min sliding window, implying that only the first two of the four 'sub-flares'  between 380 and 460\,min are sampled. The reason for this is probably that the period -- if present -- is evolving over the four cycles resulting in a periodogram peak which is too wide to be identified as a significant periodic component. This, however, does not exclude the presence of an evolving, inwards spiraling bright spot or more complicated hydrodynamic instabilities in the accretion flow around Sgr~A* \citep[e.g.][]{falanga}.

\subsection{The connection to the millimeter regime}

The Chandra X-ray Observatory and the SMA observed Sgr~A* simultaneously to this new NIR lightcurve as reported by \citet{hornstein07} and \citet{danmultwav}. Chandra started its observations at 20:00\,UT (30 July) and stopped at 08:30\,UT (31 July), so that there are simultaneous X-ray data for the entire NIR lightcurve seen in Fig.~\ref{fig1}. No X-ray flare occurred during the overlap time (23:05\,UT -- 09:00\,UT), see Fig.~6 in \citet{hornstein07}, despite the activity in the NIR with flares which are somewhat stronger than average. This may be due to the higher X-ray background caused by the steady Bondi-Hoyle accretion flow within $1\arcsec$ around Sgr~A* or/and a large surface area of the flaring region that leads to a low synchrotron self-Compton luminosity \citep[see also][]{danmultwav}. 

\begin{figure}
%\hspace{1cm}
%\begin{minipage}{8.5cm}
\includegraphics[scale=.65]{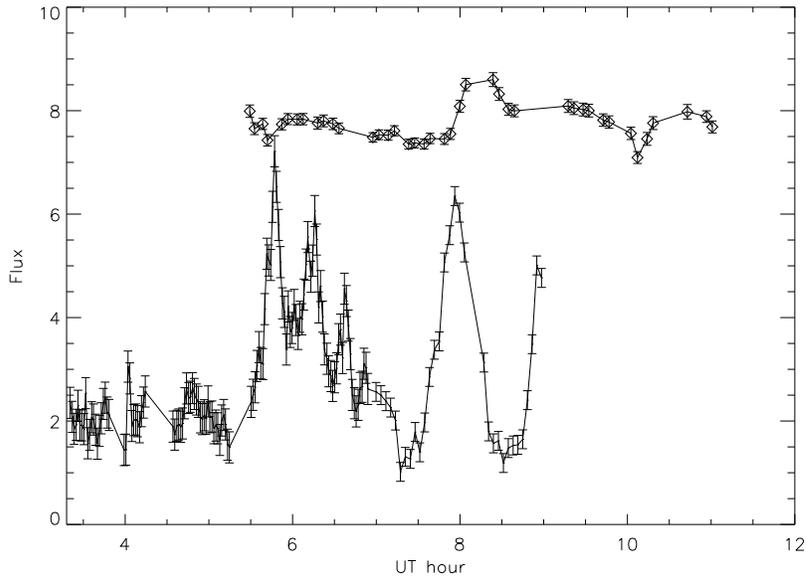}
\caption{The SMA 1.3 mm data (upper curve) from \citet{danmultwav} plotted over a common time axis with the overlapping part of the NIR lightcurve (lower curve) shown in Fig.~\ref{fig1}. The flux units are mJy for the NIR and Jy for the SMA data which have been shifted upwards by 3.5 Jy for better comparison. Please note that the first four points in the SMA lightcurve are somewhat unreliable (D. Marrone, priv. comm.). The scatter in these four points seems large compared to the error bars, which indicates that they should in fact be larger for these points than shown in the plot.}
\label{fig4}
%\end{minipage}
\end{figure}

In contrast to the X-ray lightcurve, the SMA 1.3\,mm observations, which started at 05:28\,UT, show some variability (see Fig.~\ref{fig4}). In particular, they show one flare at 08:20\,UT, roughly 20\,min after the NIR flare observed with Keck. In their analysis, \citet{danmultwav} took only the Keck NIR data into account. They interpreted this 20\,min time delay in terms of an expansion of energetic plasma similar to the model proposed by \citet{vanderlaan}. In such a model, the expanding plasma region is optically thick before, and optically thin after the peak flux. As this transition is frequency dependent, smaller and later flare peaks are expected at longer wavelengths. The 20\,min lag, however, differs from other reported time lags. In a second data set from 2006 July, \citet{danmultwav} infer a $97\pm17$\,min time lag between an X-ray flare \citep[which is supposed to be synchronous with the NIR, see][]{ecki1,ecki08} and a millimeter flare. \citet{yusef07} infer a lag of $110\pm17$\,min for the same data.

With the full NIR information at hand, the assumption of a correlation between the 08:00\,UT NIR flare and the 08:20\,UT millimeter flare becomes uncertain. Fig.~\ref{fig4} shows that there is another equally bright NIR flare preceding the 8:00\,UT flare, with no counterpart at a 20 min lag in the 1.3\,mm lightcurve. Therefore, it is equally likely that the mm flare is correlated with the preceding flare, as it is with the 8:00\,UT flare: substantially altering the interpretation of \citet{danmultwav}. 

In fact, the huge difference in the reported time lags in the 2005 mm-IR and the 2006 mm-X-ray data can be resolved with the NIR data presented here: if the $\sim$8:20\,UT mm flare (Fig.~\ref{fig4}) is correlated with the wide NIR flare at $\sim$6:00\,UT, then the inferred time lag is $\sim$140\,min. This lag is much closer to 97\,min/110\,min, than to the 20\,min assumed by \citet{danmultwav}. Note that the 8:00\,UT NIR flare might then be correlated with the SMA flare seen at 10:15\,UT.

\section{Conclusions}

In this Letter we report on a 600 minute NIR lightcurve of Sgr~A* that combines Keck II and VLT data. We showed that the Lomb-Scargle periodogram of the overall lightcurve is featureless, and is statistically consistent with a single stochastic process that has a power-law spectrum of index $-1.6$. A certain subset of the data has a prominent peak in its corresponding periodogram which, however, is not significant when analyzed in the context of the whole lightcurve. 
This points to the following dilemma in Sgr~A* research: if a periodic component exists, it is clearly a weak and transient phenomenon that persists over very few cycles, and probably has an evolving period. As we have shown, certain parts of longer pure red noise lightcurves can easily mimic such a behavior. To be able to distinguish between both scenarios, many lightcurves are needed and the range of frequencies under consideration must be narrowed down, e.g. to $0.04 - 0.07\;\mbox{min}^{-1}$ if a noticeable peak continues to occur in this range only.

We furthermore showed the difficulty with establishing the simple expanding plasmon model frequently proposed to relate NIR and mm-flares. Sgr~A* is such an active source in the NIR (and maybe not every NIR flare has a millimeter counterpart) that very long simultaneous observations are needed for a meaningful cross-correlation analysis.

\acknowledgements
We are very grateful to Dan P. Marrone for providing us with the SMA data. Some of the data presented here were obtained from Mauna Kea observatories. We are grateful to the Hawai'ian people for permitting us to study the universe from this sacred summit. This work was supported by NSF grant AST-0406816.

{\it Facilities:} \facility{VLT:Yepun (NACO)}, \facility{Keck:II (NIRC2), \facility{SMA} }

\clearpage


\end{document}